\documentclass[prb,showpacs,twocolumn,floatfix]{revtex4}
\usepackage{amsmath,amssymb}    
\usepackage{graphicx}
\usepackage{dcolumn}

\newcommand{\lco}{LiCu$_2$O$_2$}
\newcommand{\nco}{NaCu$_2$O$_2$}

\newcommand{\lcv}{LiCuVO$_4$}

\begin{document}
\title{Electronic structure, magnetic and dielectric properties of the edge-sharing copper-oxide chain compound NaCu$_{2}$O$_{2}$}
\author{Ph.~Leininger}
\affiliation{Max-Planck-Institut f\"{u}r Festk\"{o}rperforschung, Heisenbergstr. 1, D-70569 Stuttgart, Germany}
\author{M.~Rahlenbeck}
\affiliation{Max-Planck-Institut f\"{u}r Festk\"{o}rperforschung, Heisenbergstr. 1, D-70569 Stuttgart, Germany}
\author{M.~Raichle}
\affiliation{Max-Planck-Institut f\"{u}r Festk\"{o}rperforschung, Heisenbergstr. 1, D-70569 Stuttgart, Germany}
\author{B.~Bohnenbuck}
\affiliation{Max-Planck-Institut f\"{u}r Festk\"{o}rperforschung, Heisenbergstr. 1, D-70569 Stuttgart, Germany}
\author{E.~Weschke}
\affiliation{Helmholtz-Zentrum Berlin f\"{u}r Materialien und Energie, Wilhelm-Conrad-R\"{o}ntgen-Campus BESSY II, Albert-Einstein-Stra{\ss}e 15, D-12489 Berlin, Germany} 
\author{E.~Schierle}
\affiliation{Helmholtz-Zentrum Berlin f\"{u}r Materialien und Energie, Wilhelm-Conrad-R\"{o}ntgen-Campus BESSY II, Albert-Einstein-Stra{\ss}e 15, D-12489 Berlin, Germany}
\author{A.~Maljuk}
\altaffiliation[Present address: ]{Helmholtz-Zentrum Berlin f\"{u}r Materialien und Energie, D-14109 Berlin, Germany}
\affiliation{Max-Planck-Institut f\"{u}r Festk\"{o}rperforschung, Heisenbergstr. 1, D-70569 Stuttgart, Germany}
\author{C. T.~Lin}
\affiliation{Max-Planck-Institut f\"{u}r Festk\"{o}rperforschung, Heisenbergstr. 1, D-70569 Stuttgart, Germany}
\author{S.~Seki}
\affiliation{University of Tokyo, Dept. of Applied Physics, Bunkyo-ku, Tokyo 113-8656, Japan}
\author{Y.~Tokura}
\affiliation{University of Tokyo, Dept. of Applied Physics, Bunkyo-ku, Tokyo 113-8656, Japan}
\author{J.~W.~Freeland}
\affiliation{Advanced Photon Source, Argonne National Laboratory, Argonne, IL 60439, USA}
\author{B.~Keimer}
\affiliation{Max-Planck-Institut f\"{u}r Festk\"{o}rperforschung, Heisenbergstr. 1, D-70569 Stuttgart, Germany}

\date{\today}

\begin{abstract}
We report an experimental study of \nco, a Mott insulator containing chains of edge-sharing CuO$_4$ plaquettes, by polarized x-ray absorption spectroscopy (XAS), resonant magnetic x-ray scattering (RMXS), magnetic susceptibility, and pyroelectric current measurements. The XAS data show that the valence holes reside exclusively on the Cu$^{2+}$ sites within the copper-oxide spin chains and populate a $d$-orbital polarized within the CuO$_4$ plaquettes. The RMXS measurements confirm the presence of incommensurate magnetic order below a N\'eel temperature of $T_N = 11.5$ K, which was previously inferred from neutron powder diffraction and nuclear magnetic resonance data. In conjunction with the magnetic susceptibility and XAS data, they also demonstrate a new ``orbital'' selection rule for RMXS that is of general relevance for magnetic structure determinations by this technique. Dielectric property measurements reveal the absence of significant ferroelectric polarization below $T_N$, which is in striking contrast to corresponding observations on the isostructural compound \lco. The results are discussed in the context of current theories of multiferroicity.
\end{abstract}
\maketitle

\section{Introduction}

Stimulated in part by the discovery of high-temperature superconductivity, copper oxides with spin-1/2 networks ranging from zero to three dimensions have been the focus of a large amount of research during the past two decades. Among these, Mott-insulating compounds that comprise spin chains with edge-sharing CuO$_4$ plaquettes have been of particular recent interest as models of frustrated magnetism in one dimension. In these materials, the Cu-O-Cu bond angle is close to 90$^{\circ}$, so that the magnitude of the superexchange coupling between nearest-neighbor Cu$^{2+}$ spin-1/2 ions, $J_1$, is smaller than that of the next-nearest neighbor coupling, $J_2$.\cite{Mizuno1998} Depending on the sign and magnitude of the ratio $J_1/J_2$, as well as on the relative magnitudes of the spin-lattice coupling and the interchain exchange interactions, a variety of ground states have been observed in different members of this family, including spiral magnetism,\cite{Gippius2004,Capogna2005,Masuda2005,Drechsler2006,Enderle2005} collinear antiferromagnetism,\cite{Boehm1998,Fong1999} and spin-Peierls order.\cite{Hase1993} The quasi-one-dimensional spin correlations have been observed to profoundly affect charge excitations across the Mott-Hubbard gap.\cite{Matiks2009} In addition, charge \cite{Matsuda1999,Horsch2005} and a spin-density modulation \cite{Raichle2008} have been reported in materials with doped edge-sharing copper-oxide chains.

The recent discovery of ferroelectricity associated with helical magnetic order (``multiferroicity'') in the insulating edge-sharing copper-oxide chain compounds \lco\ (Ref. \onlinecite{Park2007}) and \lcv\ (Ref. \onlinecite{Naito2007}) has stimulated another wave of research on this family of materials, and it has sparked a controversy about the origin of this effect. The magneto-electric coupling is generally consistent with models according to which multiferroicity is an intrinsic consequence of spiral magnetism.\cite{Katsura2005,Sergienko2006,Mostovoy2006,Xiang2007} While some aspects of the experimental data on both compounds agree with specific predictions of these models, \cite{Seki2008,Park2007,Schrettle2008} disagreements with other predictions have also been pointed out. \cite{Park2007,Yasui2008,Xiang2007} An alternative explanation invokes orbital polarization of defects generated by inter-substitution of Li and Cu (whose ionic radii are closely similar) as the origin of the electric polarization.\cite{Moskvin2009,Moskvin2009a,Moskvin2008} This scenario is supported by the observation of substantial deviations from stoichiometry and substitutional disorder in both \lco\ and \lcv.\cite{Masuda2005,Prokofiev2004} However, the applicability of this model to \lco\ has been disputed based on the results of a recent study on single crystals with nearly stoichiometric composition.\cite{Yasui2009,Kobayashi2009}

We report the results of an experimental investigation of single crystals of \nco, a Mott insulator that is isostructural and isoelectronic to \lco\ (Fig. 1). Like \lco\ and \lcv,\cite{Gippius2004,Enderle2005,Masuda2005} \nco\ exhibits helical magnetic long-range order at low temperatures due to competing ferromagnetic nearest-neighbor and antiferromagnetic next-nearest neighbor superexchange interactions.\cite{Drechsler2006,Capogna2005} The main difference to the Li-based compounds is the large difference between the ionic radii of Na and Cu, which disfavors inter-substitution of both elements.\cite{Capogna2005,Maljuk2004} Experiments on \nco\ thus offer the chance to elucidate the intrinsic properties of this class of compounds, without interference from disorder.

This paper is organized as follows. Section II describes polarization-dependent x-ray absorption spectroscopy (XAS) experiments designed to clarify the orbital character of the valence electrons in \nco. Section III contains the results of susceptibility and resonant magnetic x-ray scattering (RMXS) experiments aimed at elucidating the magnetic properties of \nco. Section IV summarizes measurements of the electrical polarization and the dielectric constant. In the concluding Section V, the results are contrasted with observations on \lco\ and discussed in the light of current theories of multiferroicity.

\begin{figure}[t!]
\includegraphics[width=6.9 cm]{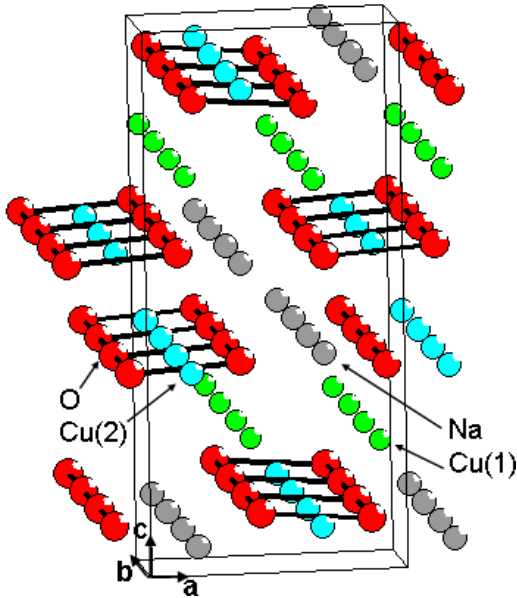}
\caption{(Color online) Sketch of the orthorhombic crystal structure of \nco\ (space group  $Pnma$). The room-temperature lattice parameters are $a$=6.2087~\AA, $b$=2.9343~\AA\ and $c$=13.0548~\AA.~\cite{Maljuk2004}}
\label{fig:structure}
\end{figure}

\section{Electronic structure}

We begin by describing XAS measurements near the Cu $L$-absorption edge aimed at determining the orbital character of the valence-electron states in \nco. The \nco\ lattice structure contains copper ions in two non-equivalent sites [Cu(1) and Cu(2) in Fig. 1]. The chemical coordination and bond lengths of these ions indicate valence states of 1+ and 2+ (with electron configurations 3$d^{10}$ and 3$d^{9}$, and spins 0 and 1/2), respectively.
However, a recent XAS study of the isostructural \lco\ had come to the conclusion that the Cu(1) ions contain an intrinsic density of holes, independent of Li-Cu inter-substitution.\cite{Chen2008} We have checked the validity of this conclusion on \nco, a compound in which the latter complication is manifestly absent.

\begin{figure}[t!]
\includegraphics[width=8.5 cm]{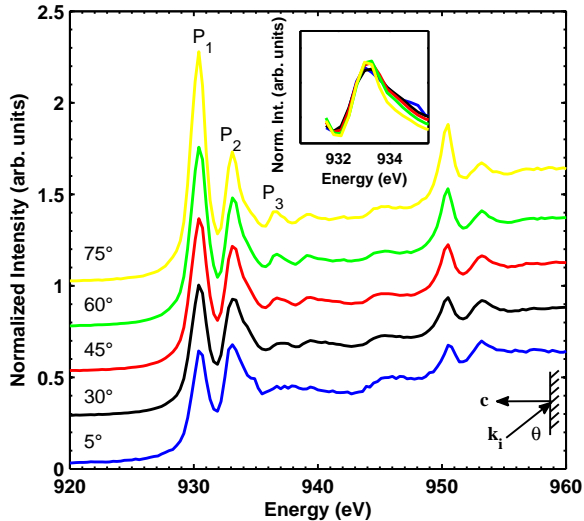}
\caption{(Color online) Polarization-dependent x-ray absorption spectra around the Cu $L_3$ and $L_2$ absorption edges of \nco. $\theta$ denotes the angle between the incident wave vector and the sample surface, which was parallel to the $ab$ plane (see lower right inset). The origins of the peaks $P_1$, $P_2$, $P_3$ are discussed in the text. The inset shows the $\theta$-dependence of the intensity of peak $P_2$ on an larger scale.}
\label{fig:xas}
\end{figure}

The experiments were performed on an untwinned, plate-like single-crystal with surface area $3 \times 2$ mm$^2$ and thickness $100~\mu\mathrm{m}$, which had been prepared by a self-flux technique already described elsewhere.\cite{Maljuk2004} The sample surface is parallel to the $ab$ plane of the orthorhombic crystal structure and hence includes the CuO$_4$ plaquettes (Fig. 1). The XAS data were taken at beamline 4-ID-C of the Advanced Photon Source in total-fluorescence-yield mode at a temperature of 150~K. The beam polarization was horizontal, and the sample was rotated such that data could be taken with different angles, $\theta$, between the incident wave vector ${\bf k_i}$ and the sample surface (Fig.~\ref{fig:xas}). All spectra were normalized to the same intensity 10~eV above the Cu $L_2$ absorption edge.

Figure 2 shows a series of XAS spectra obtained for different $\theta$. The features in the energy ranges 930-938~eV and 950-958~eV correspond to the Cu-$L_3$ and $L_2$ absorption edges, respectively. At the $L_3$ edge, three peaks are observed. The energy splitting of 2.7~eV between the first peak ($P_1$) at 930.5~eV  and the second peak ($P_2$) at 933.2~eV allows us to attribute $P_1$ to the Cu$^{2+}$ ions on site Cu(2), and $P_2$ to the Cu$^{1+}$ ions on site Cu(1).~\cite{Chen2008} The additional peak ($P_3$) at 936.7~eV is related to an excitation into non-local states.~\cite{Veenendaal1994} When ${\bf k_i}$ is turned from in-plane to out-of-plane, the intensity of $P_1$ is progressively enhanced, indicating that the valence hole on Cu$^{2+}$ occupies $d$-orbitals polarized within the CuO$_4$ plaquettes. The same polarization dependence is observed at the Cu $L_2$ absorption edge. These findings agree with density functional calculations for the structurally nearly identical \lco\ compound, which predict that the states near the Fermi level are hybrids of the Cu $d_{x^2-y^2}$ orbital and O $p_{x,y}$ orbitals in the CuO$_2$ chains.\cite{Xiang2007,Note1}

The intensity of peak $P_2$ originating from the Cu(1) ions is independent of $\theta$ within the experimental error (inset in Fig. 2), as expected for the spherically symmetric 2$p^{6}$3$d^{10}$ ground-state and 2$p^{5}$3$d^{10}$4$s^{1}$ final-state configurations in the XAS excitation of Cu$^{1+}$. We note that our data are quite different from corresponding data on \lco\ that revealed a strong polarization dependence of the $P_2$ peak intensity.~\cite{Chen2008} Based in part on electronic-structure calculations, the authors of Ref. \onlinecite{Chen2008} had attributed this observation to an intrinsic density of holes on the nominally monovalent copper ions, independent of the deviations from stoichiometry in \lco. Our data on the stoichiometric, but structurally nearly identical \nco\ do not support this interpretation. Rather, they indicate that the Cu $d$-orbitals on the Cu(1) sites of both compounds are fully occupied, as expected for the 1+ valence state, and that the polarization dependence of the corresponding XAS peak in \lco\ originates from defects created by Li-Cu inter-substitution.

\section{Magnetic properties}

Figure 3 shows the magnetic susceptibility, $\chi$, measured on an untwinned crystal from the same batch as the one described above. The broad maximum around $T$=50~K is characteristic of short-range antiferromagnetic spin correlations in low-dimensional magnets. Upon cooling below the N\'eel temperature $T_N$=11.5~K, $\chi$ exhibits a sharp drop for magnetic fields, \textbf{H}, along $b$ or $c$, while a slight upturn is observed for $\textbf{H}\parallel a$. This implies that the direction of the magnetic moment in the N\'eel state is predominantly in the $bc$ plane, consistent with a $bc$-polarized spiral inferred from neutron powder diffraction and nuclear magnetic resonance (NMR) measurements.\cite{Capogna2005,Drechsler2006,Note2} A previously reported \cite{Capogna2005,Drechsler2006} anomaly in $\chi (T)$ for $T \sim 8$ K was not observed in our samples. Measurements of $\chi$(\textbf{H}) up to 7 T (not shown) did not reveal any anomalies indicative of spin-flop transitions.

\begin{figure}[t!]
\includegraphics[width=8.5 cm]{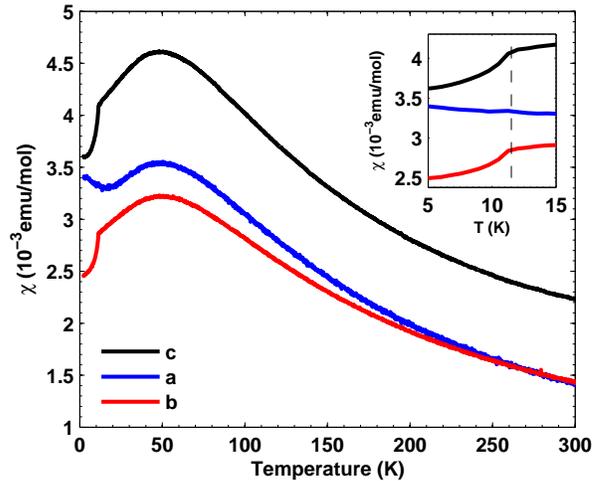}
\caption{(Color online) Temperature dependence of the magnetic susceptibility along the principal crystallographic axes measured with a magnetic field H=1000~Oe. The inset highlights the data at low temperature.}
\label{fig:chi}
\end{figure}

RMXS measurements with photon energies near the Cu $L_3$ edge were performed in order to further investigate the spin correlations below $T_N$. The experiments were performed at the U46-PGM1 beamline at the BESSY II synchrotron in Berlin, Germany, using a two-circle diffractometer with horizontal scattering geometry designed at the Freie Universit\"{a}t Berlin. Silver paste was used to glue the crystal onto a copper goniometer attached to a continuous flow cryostat, such that the $a$- and $b$-axes were parallel to the diffraction plane. A $800~\mathrm{nm}$ thick aluminum foil provided sufficient heat shielding to reach sample temperatures as low as 3~K. The incident beam was polarized either horizontally (H) or vertically (V), {\it i.e.} parallel or perpendicular to the scattering plane. The diffracted beam was detected without polarization analysis.

\begin{figure}[t!]
\includegraphics[width=8.5 cm]{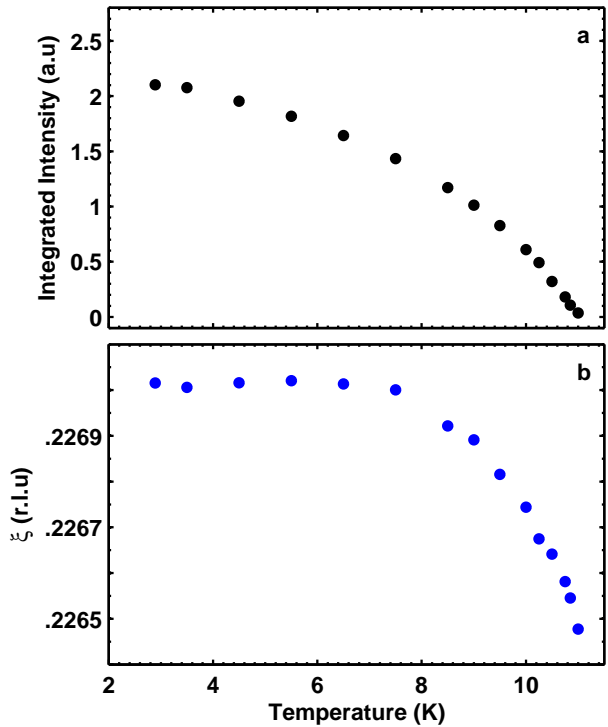}
\caption{(Color online) (a) Temperature dependence of the intensity of the magnetic Bragg reflection at \textbf{Q}=(0.5 $\xi$ 0), extracted from fits to RMXS data. (b) $b$-axis component of \textbf{Q} as a function of temperature.}
\label{fig:tdpce}
\end{figure}

At temperatures below $T_N$, a resonant diffraction peak was observed at wave vector transfer \textbf{Q} = (0.5 $\xi$ 0), with incommensurate component $\xi \sim 0.227$ along the spin chains. (The components of \textbf{Q} are indexed based on the orthorhombic crystal structure; see Fig. 1.) The origin of the peak was identified as magnetic based on the temperature dependence of its intensity (Fig. 4a). The RMXS data thus confirm the propagation vector extracted from powder neutron diffraction measurements.\cite{Capogna2005} The width of the diffraction peak, measured along the propagation vector, is temperature independent and yields a lower bound of 3500 \AA\ on the magnetic domain size in the $ab$ plane. The temperature variation of the incommensurate wave vector below $T_N$ (Fig. 4b) is much smaller than the corresponding variation in \lco, where a two-step transition to magnetic long-range order was observed. \cite{Rusydi2008,Huang2008} The energy dependence of the (0.5 $\xi$ 0) peak intensity (Fig. 5) shows two minima at 931.6 and 934.3~eV, which can be attributed to strong absorption at the  $L_3$ edges of Cu$^{2+}$ and Cu$^{+}$ (dotted lines in Fig. 5). Similar data were reported for \lco\ (Ref.~\onlinecite{Huang2008a}). The resonant magnetic Bragg reflection was also observed when the photon energy was tuned to the Cu $L_2$-absorption edge of \nco, but an equivalent experiment at the Cu $K$-edge yielded a null result. This is consistent with the expectation that the resonant enhancement of the RMXS cross section at transition-metal $L$-edges is much larger than at $K$-edges.

We now discuss the dependence of the RMXS cross section on the incident beam polarization $\hat{\epsilon}_i$ and on the azimuthal angle $\Psi$. The scattering geometry and the definition of $\Psi$ are illustrated in the inset of Fig. 6. Note that \textbf{Q} of the magnetic Bragg reflection approximately bisects the angle between the $a$- and $b$-axes, and that for $\Psi = 0^\circ$ the final wave vector $\mathbf{k}_f$ is nearly along $\hat{b}$ (and the outgoing polarization $\hat{\epsilon}_f$ is hence nearly $\perp \hat{b}$). For $\Psi = 180^\circ$, on the other hand, $\mathbf{k}_f$ is nearly along $\hat{a}$ ($\hat{\epsilon}_f$ nearly $\perp \hat{a}$). According to the standard treatment of RMXS,\cite{Hill1996} the magnetic scattering amplitude is proportional to [$(\hat{\epsilon}_f\times \hat{\epsilon}_i)\cdot \mathbf{m}$], where $\mathbf{m}$ is the magnetic moment. The susceptibility indicates that $\mathbf{m}$ is confined to the $bc$ plane below $T_N$ (Fig. 3).
For $\Psi = 0^\circ$ and V (H) incident polarization, one would hence expect the scattering amplitude to be mainly proportional to $m_b$ ($m_c$), while for $\Psi = 180^\circ$ it should nearly vanish (be sensitive to both $m_b$ and $m_c$). Since the susceptibility further indicates comparable magnitudes of $m_b$ and $m_c$, the ratio $I_V/I_H$ is expected to be of the order of unity for $\Psi = 0^\circ$, and to nearly vanish for $\Psi = 180^\circ$.
\begin{figure}[t!]
\includegraphics[width=8.5 cm]{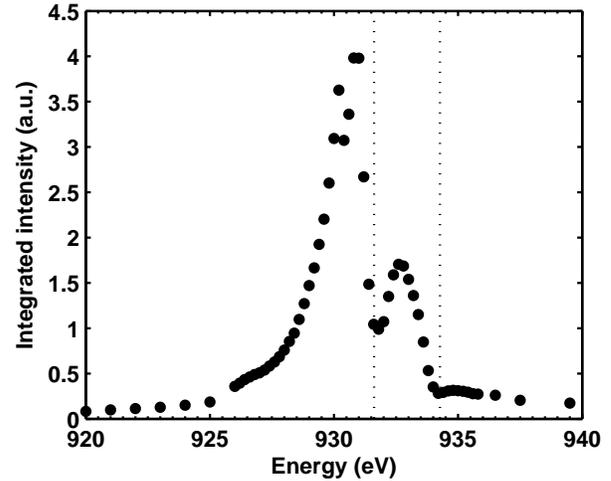}
\caption{(Color online) Energy dependence of the scattered intensity at the (0.5 0.227 0) magnetic reflection near the Cu $L_3$ absorption edge. The data were taken at $T = 3$ K. The dotted lines at 931.6~eV and 934.3~eV represent the energies of the Cu$^{2+}$ and Cu$^{1+}$ absorption edges, which were extracted from the widths of longitudinal momentum scans through the magnetic reflection.}
 \label{fig:edpce}
\end{figure}

\begin{figure}[t!]
\includegraphics[width=8.5 cm]{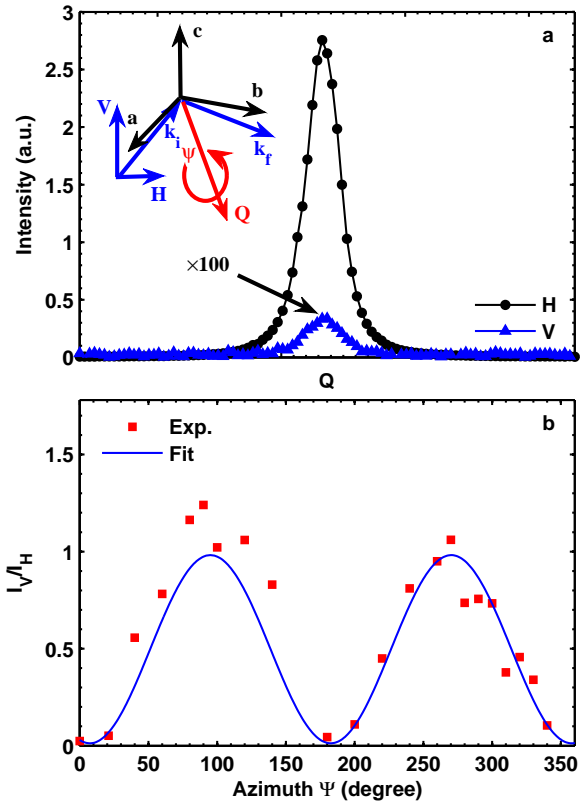}
\caption{(Color online) (a) Longitudinal reciprocal-space scans through the \textbf{Q}=(0.5 $\xi$ 0) reflection taken with horizontal (H) and vertical (V) polarization of the incident x-ray beam. The data for V-polarization were multiplied by 100. (b) Azimuthal-angle ($\Psi$) dependence of the ratio of Bragg peak intensities measured with V- and H-polarized incident beams. The line is the result of a fit to the function $A \sin^2 \Psi$, as discussed in the text. A small offset in $\Psi$ was also included in the fit in order to account for imperfect sample alignment. The inset illustrates the definition of $\Psi$ and the scattering geometry of $\Psi = 0^\circ$.}
\end{figure}

In striking contrast to this expectation, we have found that $I_V$
is about three orders of magnitude smaller than $I_H$ for $\Psi = 0^\circ$
(Fig. 6a). Moreover, the azimuthal-angle dependence of the ratio
$I_V/I_H$ (which should not be strongly affected by distortions arising
from sample-size or surface effects) is well described by the function
$A \sin^2\Psi$ with $A \sim 1$ (Fig. 6b), which would imply $m_a \sim m_b
\sim 0$ according to the standard description of
RMXS.\cite{Hill1996} A recent generalization of the theory of RMXS
that takes the symmetry of the valence electron wave function into
account\cite{Haverkort2009} resolves the apparent
contradiction between the susceptibility and RMXS data.
In the case at hand (which is explicitly treated in Ref.
\onlinecite{Haverkort2009}), it was shown that a selection rule
precludes excitations of $2p$ core electrons into the partially
occupied, planar $3d_{x^2-y^2}$ orbital of Cu$^{2+}$ (see Section II) by
vertically polarized photons. This implies that $I_V = 0$ for both
$\Psi = 0^\circ$ and $180^\circ$. Our data thus confirm the revised
analysis of Ref. \onlinecite{Haverkort2009} and highlight its
importance for magnetic structure determination by RMXS.

\section{Dielectric properties}

Measurements of the electrical polarization $\textbf{P}$ and the dielectric constant $\varepsilon$ were performed on a \nco\ crystal with dimensions $\approx 5 \times 2 \times 0.4$ mm$^3$. Due to the small thickness of the sample, the experiments were restricted to the $c$-axis components of both quantities. Electrodes were fabricated by covering the $ab$ faces of the crystal by silver paste. 

The sample was loaded into a 14 T cryomagnet, and cooled down to 2 K while applying a poling electric field of 250 kV/m. Then, $\mathrm{P}_c$ was determined through a time-integral of the pyroelectric current during a temperature sweep rate of 20 K/min. $\varepsilon_c$ was measured with an $LCR$ meter operating at a frequency of 100 kHz with a temperature sweep rate of +2 K/min.

\begin{figure}[t!]
\includegraphics[width=8.5 cm]{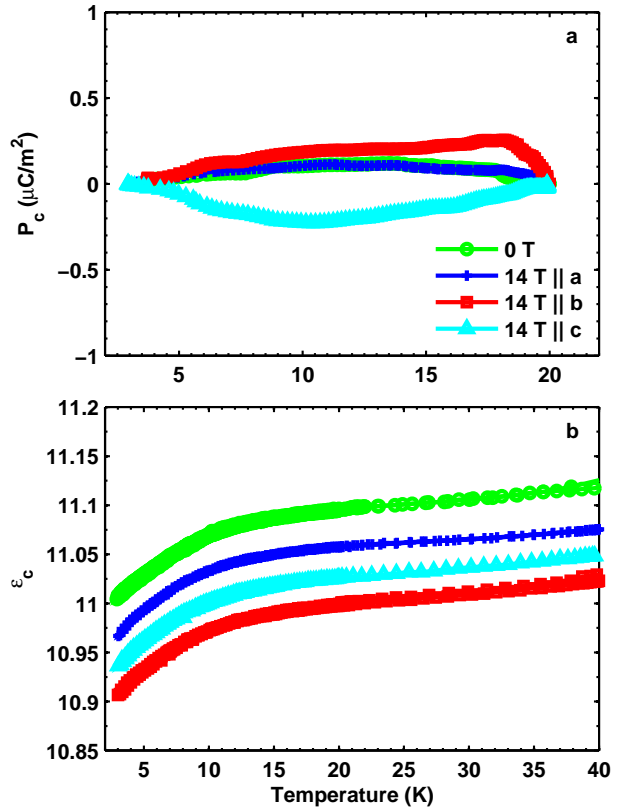}
\caption{(Color online) (a) Electrical polarization and (b) dielectric constant measured along the $c$-axis for $\textbf{H} = 0$ and 14 T applied along the three principal crystallographic directions. The data in panel (b) were arbitrarily shifted in vertical by 
$<$ 5\% for clarity and for taking into account small changes in the electrodes connection.} 
\end{figure}

As shown in Fig. 7a, neither a spontaneous nor a magnetic-field-induced electric polarization were detected for magnetic fields $\textbf{H}\leq 14$ T within the experimental sensitivity of $0.3 \,\mu$C/m$^2$. This upper bound is more than an order of magnitude lower than the $\mathrm{P}_c$ observed in \lco\ for $T \ll T_N$.~\cite{Park2007,Seki2008} The dielectric constant exhibits a subtle downturn upon cooling below $T_N$ (Fig. 7b). Effects of this magnitude have also been observed in other antiferromagnetic insulators~\cite{Katsufuji2001} and may be a consequence of conventional spin-phonon coupling. Sharp spikes in the temperature dependence of $\varepsilon_c$ such as the one observed at the N\'eel temperature of \lco\ (Refs. \onlinecite{Park2007, Seki2008}) are not detected within the experimental sensitivity.

\section{Conclusions}

At first sight, the absence of multiferroicity in \nco\ appears to contradict models that predict a ferroelectric polarization of the form $\mathbf{P} \propto \hat{n}_{ij} \times ( \mathbf{S}_{i} \times \mathbf{S}_{j})$ (where $\hat{n}_{i,j}$ is a vector connecting nearest-neighbor spins $\textbf{S}_{i,j}$) as an intrinsic consequence of spiral magnetism.\cite{Katsura2005,Sergienko2006,Mostovoy2006,Xiang2007} Neutron powder diffraction and NMR measurements have indicated a $bc$-polarized spiral with propagation vector along $b$,\cite{Drechsler2006,Capogna2005} which is also consistent with the susceptibility and RMXS data presented here. The models thus predict a ferroelectric polarization along $c$, which was not observed (Fig. 7). Refinement of a comprehensive set of single-crystal magnetic neutron diffraction data has, however, revealed a more complex magnetic structure in which the spin chains within the unit cell (Fig. 1) exhibit spiral states with opposite helicities\cite{Capogna2009} so that their ferroelectric polarizations (if present) are expected to cancel out. Our data are consistent with such an antiferroelectric state below $T_N$, and hence do not contradict the theoretical work of Refs. \onlinecite{Katsura2005,Sergienko2006,Mostovoy2006}.

We have also shown that the Cu(1) sites of the \nco\ structure exhibit a full-shell configuration and are hence electronically inactive. This agrees with prior considerations based on the chemical coordination of these sites, but disagrees with the conclusions of a recent XAS study of \lco\ according to which they exhibit an intrinsic density of holes.\cite{Chen2008} The orbital polarization detected in the Cu(1) XAS spectra of \lco\ is thus presumably a consequence of defects generated by Li-Cu inter-substitution. In view of the nearly identical lattice structures of \lco\ and \nco, it seems natural to associate these defects with the ferroelectric polarization detected in \lco, but not in \nco. We cannot rule out, however, that the qualitative difference between the dielectric properties of both compounds originates in a subtle difference between their magnetic structures.

Finally, the simultaneous determination of the orbital occupation of the valence electrons of \nco\ by XAS and their magnetic structure by RXMS allowed us to experimentally confirm an ``orbital'' selection rule for RMXS recently proposed on the basis of a theoretical analysis.\cite{Haverkort2009} Especially for non-collinear spin structures, it is thus important to consider the orbital occupation before drawing conclusions about the spin polarization from RXMS data.

\section*{ACKNOWLEDGMENTS}
We thank E. Dudzik and R. Feyerherm for assistance during the experiment at the 7T-MPW-MagS beamline of BESSY. We thank P. Horsch, R. K. Kremer, D. Efremov and M. W. Haverkort for useful discussions. Work at the Advanced Photon Source, Argonne is supported by the U.S. Department of Energy, Office of Science under Contract No. DEAC02-06CH11357. We thank C.~Sch\"{u}{\ss}ler-Langeheine and the SFB 608 for instrumental support.

\end{document}